\documentclass[11pt]{article}
\usepackage{amsthm,amsmath,amssymb,bm,color,natbib,multirow}
\usepackage[hypertexnames=false,bookmarksnumbered]{hyperref}
\hypersetup{colorlinks=true,citecolor=black}
\usepackage{graphicx}

\theoremstyle{plain}
\newtheorem{thm}{Theorem}%[section]
\newtheorem{lemma}{Lemma}%[section]
\newcommand{\bd}{{\boldsymbol{d}}}

\newcommand{\bu}{{\boldsymbol{u}}}

\newcommand{\bx}{{\boldsymbol{x}}}
\newcommand{\by}{{\boldsymbol{y}}}
\newcommand{\bz}{{\boldsymbol{z}}}

\newcommand{\bA}{{\boldsymbol{A}}}

\newcommand{\bG}{{\boldsymbol{G}}}
\newcommand{\bH}{{\boldsymbol{H}}}
\newcommand{\bI}{{\boldsymbol{I}}}

\newcommand{\bU}{{\boldsymbol{U}}}

\newcommand{\bW}{{\boldsymbol{W}}}
\newcommand{\bZ}{{\boldsymbol{Z}}}
\newcommand{\bY}{{\boldsymbol{Y}}}
\newcommand{\bX}{{\boldsymbol{X}}}

\newcommand{\bone}{{\boldsymbol{1}}}

\newcommand{\bmu}{{\boldsymbol{\mu}}}
\newcommand{\bbeta}{{\boldsymbol{\beta}}}
\newcommand{\btheta}{{\boldsymbol{\theta}}}
\newcommand{\besp}{{\boldsymbol{\varepsilon}}}
\newcommand{\bTheta}{{\boldsymbol{\Theta}}}

\newcommand{\newstuff}[1]{\textcolor{black}{#1}}
\newcommand{\ignore}[1]{}

\newcommand{\brm}[1]{\bm{\mathrm{#1}}}

\newcommand{\RSS}{\text{RSS}}

\title{\vspace*{-2cm}Uncertainty Quantification for High Dimensional Sparse Nonparametric Additive Models}
\author{
Qi Gao\thanks{Department of Statistics, University of California at Davis, 4118 Mathematical Sciences Building, One Shields Avenue, Davis, CA 95616, USA. Email: {\ttfamily qigao@ucdavis.edu}}
\and
Randy C. S. Lai\thanks{Department of Mathematics and Statistics, University of Maine, 5752 Neville Hall, Room 333, Orono, ME 04469, USA. Email: {\ttfamily chushing.lai@maine.edu}}
\and
Thomas C. M. Lee\thanks{Corresponding author.  Department of Statistics, University of California at Davis, 4118 Mathematical Sciences Building, One Shields Avenue, Davis, CA 95616, USA. Email: {\ttfamily tcmlee@ucdavis.edu}}
\and
Yao Li\thanks{Department of Statistics, University of California at Davis, 4118 Mathematical Sciences Building, One Shields Avenue, Davis, CA 95616, USA. Email: {\ttfamily yaoli@ucdavis.edu}} \\
}
\date{June 24, 2018; revised June 28, 2019}

\begin{document}
\maketitle

\vspace*{-1cm}
\begin{abstract}

Statistical inference in high dimensional settings has recently attracted enormous attention within the literature.  However, most published work focuses on the parametric linear regression problem.  This paper considers an important extension of this problem: statistical inference for high dimensional sparse nonparametric additive models.  To be more precise, this paper develops a methodology for constructing a probability density function on the set of all candidate models.  This methodology can also be applied to construct confidence intervals for various quantities of interest (such as noise variance) and confidence bands for the additive functions.  This methodology is derived using a generalized fiducial inference framework.  It is shown that results produced by the proposed methodology enjoy correct asymptotic frequentist properties.  Empirical results obtained from numerical experimentation verify this theoretical claim.  Lastly, the methodology is applied to a gene expression data set and discovers new findings for which most existing methods based on parametric linear modeling failed to observe.

\end{abstract}

\section{Introduction}

Nonparametric additive models, given their flexibility, have long been a popular tool for studying
the effects of covariates in regression problems \citep[e.g.,][]{Friedman81, Stone85}.  Given a set of $n$ independently and identically distributed observations $\{(Y_i, \bX_i)\}_{i=1}^n$, with $Y_i$ being the $i$-th response and $\bX_i=(X_{i1}, \ldots, X_{ip})^\top$ as the $i$-th $p$-dimensional covariate, a nonparametric additive model is defined as
\begin{align}
Y_i = \mu + \sum_{j=1}^{p} f_j(X_{ij}) + \varepsilon_i, \quad i=1, \ldots, n,
\label{eq:additive_model}
\end{align}
where $\mu$ is an intercept term, the $f_j$'s are unknown (and usually smooth) functions, and $\varepsilon_i$ is an independent random error with mean zero and finite variance $\sigma^2$.  Here this paper allows the possibility that $p$ is greater than $n$, which implies some of the functions $f_j$'s are zero.

There is a rich literature on the estimation of the functions $f_j$'s in~(\ref{eq:additive_model}) when $p<n$ is fixed.  For example, \cite{Stone85} develops spline estimators that achieve the same optimal rate of convergence for general $p$ as for $p=1$ under some assumptions. \cite{Buja-et-al89} propose a backfitting algorithm to estimate the functions with linear smoothers and prove its convergence.  For fixed $p$ and under some mild regularity conditions,
%\cite{Horowitz-and-Mammen} and
\cite{Horowitz-et-al06} obtain oracle efficient estimators using a two-step procedure which are asymptotically normal with convergence rate $n^{-2/5}$ in probability.

In high dimensional settings where $p>n$, much work has also been done in variable selection; i.e.,
selecting (and estimating) the significant $f_j$'s. \cite{Meier-et-al09} propose using a new sparsity-smoothness penalty for variable selection and provide oracle results which lead to asymptotic optimality of their estimator for high dimensional sparse additive models. \cite{Ravikumar-et-al09} derive a sparse backfitting algorithm for variable selection with a penalty based on the $l_2$ norm of the mean value of the nonparametric components. Their algorithm decouples smoothing and sparsity and is applicable to any nonparametric smoother.  \cite{Huang-et-al10} apply adaptive group Lasso to select significant $f_j$'s and provide conditions for achieving selection consistency.

In recent years there has been a growing body of work in statistical inference for high dimensional linear parametric models.  For example, \cite{buhlmann2013}, \cite{javanmard2014}, \cite{vanet-al2014} and \cite{zhang2014} study hypothesis testing and confidence intervals for low dimensional parameters in high dimensional linear and generalized linear models.  Their approaches are mostly based on ``de-biasing'' or ``de-sparsifying'' a regularized regression estimator such as Lasso.  \cite{chatterjee2013} and \cite{lopes2014} examine properties of the residual bootstrap for high dimensional regression.  \cite{lee-et-al2016} and \cite{Tibshirani-et-al16} consider the exact post-selection inference for sequential regression procedures conditioning on the selected models.  Lastly, the empirical Bayes approach has also been adopted; e.g., see \cite{martin2017empirical}.

However, much less attention is given to statistical inference for nonparametric additive models, especially in high dimensional settings.  \cite{Fan-Jiang05} extend the generalized likelihood ratio tests to additive models estimated by backfitting to determine if a specific additive component is significant or admits a certain parametric form.  However, these authors do not consider the cases where $p>n$ and inferences for some parameters such as $\sigma$.  More recently \cite{Lu-et-al15} propose two types of confidence bands for the marginal influence function in a novel high dimensional nonparametric model, termed ATLAS, which is a generalization of the sparse additive model, although no inference procedure is provided for other model components.  Lastly, various Bayesian methods have been also proposed, including \citet{Scheipl-et-al12} and \citet{Shang-Li14}.  However, none of these methods is designed to provide uncertainty quantification for high dimensional nonparametric additive models.

The main goal of this paper is to address the inference problem for high dimensional nonparametric additive models.  To be more specific, this paper develops a method that quantifies the uncertainties in the estimated parameters and selected models.  This method is based on the generalized fiducial inference (GFI) framework \citep{Hannig-et-al16}, which has been shown to possess extremely good properties, both theoretical and empirical, in various inference problems.  To the best of our knowledge, this is the first time that uncertainty quantification is being formally considered for high dimensional additive models.
%To the best of our knowledge, this is the first time that such problems are formally considered.

The remainder of this paper proceeds as follows. In the next section, we first present a spline representation of nonparametric additive models upon which our inference will be based.  In Section~\ref{sec:method} we introduce the GFI framework and formally describe our proposed inference method for sparse and high dimensional nonparametric additive models.  Section~\ref{sec:theory} examines the theoretical properties of the proposed method while Section~\ref{sec:numerical} illustrates its empirical properties via numerical experiments and a real data example.  Lastly, concluding remarks are offered in Section~\ref{sec:conclude} while proofs of theoretical results are delayed in the appendix.

\section{Spline Modeling of Additive Functions}
\label{sec:spline}
The functions $f_j$'s in nonparametric additive models are commonly modeled by splines $f_{nj}$'s in practice.  A spline function is a piecewise polynomial function, usually cubic, that is connected together at knots.  Here we state the standard conditions and definition for spline functions following, for example, \cite{Stone85} and \cite{Huang-et-al10}.

Suppose that $X_j \in \mathcal{X}_j$ where $\mathcal{X}_j=[a,b]$ for finite numbers $a<b$ and $E(Y^2)<\infty$. To ensure identifiability, we assume $Ef_j(X_j)=0$ for $j=1,\dots,p$. Let $K$ be the number of knots for a partition of $[a,b]$ satisfying condition (A2) stated in Section~\ref{sec:theory} below.  Let $\mathcal{S}_n$ be the collection of functions $s$ on $[a,b]$ satisfying the following two conditions: (i) $s$ is a polynomial of degree $l$ (or less) on each sub-interval, and, (ii) for two integers $l$ and $l'$ satisfying $l\geq 2$ and $0\leq l'<l-1$,  $s$ is $l'$-times continuously differentiable on $[a,b]$.
%A function satisfying both (i) and (ii) is called a spline and $\mathcal{S}_n$ is the space of polynomial splines of degree $l$.

Then there exists a normalized B-spline basis $\{\varphi_{k}(\cdot),
k=1,\ldots,h_n\}$, $h_n=K+l$ for $\mathcal{S}_n$, such that for any $f_{nj}\in \mathcal{S}_n$,
\begin{align}
f_{nj}(x) = \sum_{k=1}^{h_n} \beta_{jk} \varphi_{jk}(x),
\label{eq:spline_rep}
\end{align}
where $\beta_{jk}$ is the coefficient of the basis function $\varphi_{jk}(x)$, $k=1,\ldots,h_n$. As shown in Lemma~\ref{spline} below, $f_j$'s can be well approximated by functions in $\mathcal{S}_n$ under certain smoothness conditions.  Thus in the rest of this paper, for the purpose of expediting technical calculations, we shall assume that the spline representation is exact for the additive functions $f_j$'s.

In matrix notation, equation~\eqref{eq:additive_model} can be rewritten in the following form
\begin{align}
\label{eq:spline_matrix}
\bY = \mu \bone + \bZ \bbeta + \besp,
\end{align}
where $\bY = (Y_1,\ldots,Y_n)^\top$, $\bZ$ is a $n\times(h_np)$ matrix with $i$th row equals
to $(\varphi_{11}(X_{i1}), \varphi_{12}(X_{i1}), \ldots, \break \varphi_{1h_n}(X_{i1}) ,\ldots,
\varphi_{p1}(X_{ip}), \varphi_{p2}(X_{ip}), \ldots, \varphi_{ph_n}(X_{ip}) )$, $\bbeta =
(\beta_{11},\ldots, \beta_{1h_n},\ldots,\beta_{p1},\ldots, \beta_{p h_n})^\top$ and $\besp =
(\varepsilon_1,\ldots,\varepsilon_n)$. This linear representation of additive models provides us a proxy to apply the GFI methodology on high dimensional regression models as described in \cite{Lai-et-al15}.

\section{Methodology}
\label{sec:method}
\subsection{Generalized Fiducial Inference}

The original idea of fiducial inference can be dated back to the 1930's. \cite{fisher1930inverse} introduces fiducial inference as an alternative to Bayesian procedures with the goal to assign an appropriate statistical distribution on the parameters of a parametric family of distributions.  One well-known criticism of the classical Bayesian procedures is the need of specifying prior distributions for the parameters.  Fisher's proposal aims to avoid such an issue by considering a switching mechanism between the parameters and the observations, in a way very similar to the procedure of obtaining parameter estimates by maximizing the likelihood function.  In spite of Fisher's continuous effort in establishing a formal inference framework via the fiducial argument, it has been overlooked for many years by the majority of the statistics community.  Interested readers are referred to \cite{Hannig-et-al16} where a detailed discussion about the history of fiducial inference and numerous related references can be found.

In recent years, there has been increasing interest in reformulating the somewhat abandoned fiducial concepts. These modern modifications include Dempster-Shafer theory \citep{dempster2008dempster}, its relative inferential models \citep{martin2010dempster,martin2013inferential,martin2015inferential} and confidence distribution \citep{xie2013confidence}. One such modern formulation of Fisher's fiducial inference is the so-called generalized fiducial inference or GFI \citep{hannig2009b,Hannig-et-al16}.  GFI has been applied successfully in many classical and modern problems, including wavelet regression \citep{hannig2009c}, linear mixed models \citep{cisewski2012generalized} and logistic regression model \citep{liu2016generalized}.  In particular, \cite{Lai-et-al15} successfully apply GFI on ultra-high dimension regression models and show that the resulting GFI inference procedure has excellent theoretical and practical performance. %Inspired by their work, we are interested in studying a generalization of their model to the nonparametric additive models.

\subsection{A Recipe for Applying GFI}
The most significant idea behind the philosophy of GFI is a switching principle. It begins by realizing that any $n$-dimensional observation $\bY$ can be viewed as an outcome of an equation:
\begin{align}
\bY = \bG(\btheta, \bU),
\end{align}
where $\btheta\in\bTheta$ is a $p$-dimensional fixed parameter vector which determines the distribution of $\bY$, $\bU$ is a random variable whose distribution is known and does not depend on $\btheta$, and $\bG$ is a parametric deterministic function relating $\bY$ and $\btheta$.  Such a relationship is sometimes known as a ``structural equation'' in other areas of study.  There may be more than one structural equation for any given distribution of a random vector $\bY$. If the elements of $\bY$ are independent, a naive choice of $\bG$ would be the inverse distribution function for each element and $\bU$ would be just an i.i.d. uniformly $(0,1)$ random vector.

The switching principle states that, if $\bY = \by$ is observed, a distribution of $\btheta$ can be defined by inverting the relationship of $\by$ and $\btheta$ while continuing to believe that the same relation holds and the distribution of $\bU$ remains unchanged.  With this thinking, for any $\by$, one could define the set $\left\{\btheta: \by = \bG(\btheta, \bU^*)\right\}$ as the inverse mapping of $\bG$ and $\bU^*$ is distributed identically as $\bU$. This random set could be empty if there are no $\btheta$'s such that $\by = \bG(\btheta, \bU^*)$, or it could have more than one element if there is more than one $\btheta$ such that $\by = \bG(\btheta, \bU^*)$.  The support of $\bU^*$ could be renormalized to assure that there is at least one solution of the equation. For those values of $\bU^*$ resulting in multiple solutions, \cite{hannig2009b} suggested randomly picking an element from the random set $\left\{\btheta: \by = \bG(\btheta, \bU^*)\right\}$.

This algorithm yields a random sample of $\btheta$ if $\bU^*$ is repeatedly sampled.  The resulting random sample of $\btheta$ is called a fiducial sample of $\btheta$, on which statistical inferences of $\btheta$ could be based.  The density function of $\btheta$ is also implicitly defined via this algorithm and is denoted as $r(\btheta|\by)$. The function $r(\btheta|\by)$ is called the generalized fiducial density and \cite{Hannig-et-al16} show that, under reasonable smoothness assumptions of the likelihood function of $\bY$, a version of the generalized fiducial density is given by
\begin{align}
\label{eq:generalized_fiducial_density}
    r(\btheta|\by)=\frac{f(\by,\btheta)J(\by,\btheta)}{\int_{\bTheta}f(\by,\btheta')J(\by,\btheta')d\btheta'},
\end{align}
where
$$
J(\by,\btheta)=D\left( \frac{\bd}{\bd\btheta}\bG(\btheta,\bu)|_{\bu=\bG^{-1}(\by,\btheta)}\right),
$$
$D(\bA)=(\det \bA ^\mathsf{T} \bA)^{1/2}$ and $\bu=\bG^{-1}(\by,\btheta)$ is the value of $\bu$ such that $\by = \bG(\btheta, \bu)$.

Although the generalized fiducial density in equation~\eqref{eq:generalized_fiducial_density}
provides an explicit expression for the distribution of $\btheta$, it is not always possible to
calculate its form analytically.  For example, it is very often that $r(\btheta|\by)$ is known
only up to a normalizing constant, and in such cases one may need to use Monte Carlo techniques to
simulate a fiducial sample.  Besides conventional Monte Carlo techniques,
\cite{hannig2014computational} consider a non-intrusive method for models for which closed form
densities are not available.

Model selection was introduced into the GFI paradigm by \cite{hannig2009c} in the context of wavelet
regression.  The most significant challenge is to incorporate the uncertainty due to model selection into the problem setup. To facilitate the notation, denote now the structural equation of a particular model $M$ as
\begin{align}
\bY = \bG(M, \btheta_M, \bU), \quad M \in \mathcal M,
\label{eqn:G}
\end{align}
where $\mathcal M$ is a collection of models. Thus, for any given model, equation~
\eqref{eq:generalized_fiducial_density} gives the corresponding generalized fiducial density for $\btheta$, which is now represented as $r(\btheta | \by, M)$.  As stated by \cite{Hannig-et-al16}, similar to MLE, GFI tends to favor large models, therefore additional penalty and assumptions about the model size are needed to account for the model complexity.  These authors also argue for introducing penalty in the GFI framework which leads to the following marginal generalized fiducial probability $r(M)$ of model $M$:
\begin{equation}
    \label{eq:gfd_model}
    r(M) = \frac{\int r(\btheta | \by, M) q^{|M|} d \btheta_M}
{\sum_{M' \in \mathcal M}\int r(\btheta | \by, M') q^{|M'|} d \btheta_{M'}},
\end{equation}
where $q$ is a constant determined by the penalty and $|M|$ is the number of parameters of the model
$M$.  Note that for brevity we suppress the dependence of $\by$ in the notation of $r(M)$.  The value of $q$ can be interpreted as the prior sparsity rate of the predictors under the Bayesian framework, or can be viewed as a solely penalty term as in the context of frequentists. In GFI, $q$ can be thought as the probability of observing a structural equation for a specific predictor.  For the $p<n$ scenario, one can choose $q$ as $n^{-1/2}$ which results in the classical BIC penalty.  However, for the more general and high dimensional setting, the choice of $q$ will need to be adjusted.  One possibility is to set $q \propto p^{-1}$ which matches the extended Bayesian information criterion (EBIC) of \citet{Luo-Chen13} with $\gamma=1$, where $\gamma$ is a user-specified parameter for EBIC.  Such a choice of $q$ is justified by the theoretical results to be presented below.  Throughout all our numerical work, we set $q=0.2 p^{-1}$.

\subsection{GFI for Nonparametric Additive Models}
% \subsection{Generalized fiducial distribution of the coefficients}

This subsection applies the above results to nonparametric additive models and obtains the corresponding generalized fiducial probability.  Without loss of generality, first assume that in~(\ref{eq:spline_matrix}) $\mu=0$ and the random error $\besp$ is normally distributed with covariance $\text{diag}(\sigma^2,\ldots,\sigma^2)$. Let $M$ denote any candidate model, $M_0$ be the true model and $\bH$ be the projection matrix of $\bZ$; i.e., $\bH=\bZ(\bZ^T\bZ)^{-1}\bZ^T$. The residual sum of squares RSS is given by $\RSS=\|\by-\bH\by\|^2$.
\newstuff{
 The structural equation~(\ref{eqn:G}) is now
\begin{align}
\bY = \bG(M, \btheta_M, \bU) = \bZ \bbeta + \sigma \bU, \quad M \in \mathcal M.
\end{align}
}
It can be shown that for the parameters $\btheta = (\sigma, \bbeta)^\top$ in model~(\ref{eq:spline_matrix}) (with $\mu=0$) \citep[e.g.,][]{Lai-et-al15}
$$J(\by,\btheta)=\sigma^{-1}|\text{det}(\bZ'\bZ)|^{1/2}\RSS^{1/2}.$$
Therefore the generalized fiducial density of $\btheta$ given any model $M$ is
\begin{equation}
    r(\btheta | \by, M) =  \frac{ \sigma^{-1} \left[\det(\bZ^\top \bZ)\right]^{1/2} \RSS^{1/2}
    \left(\frac {1} {2 \pi
    \sigma^2}\right)^{n/2} \exp\left\{-\frac{1}
    {2\sigma^2} (\by - \bZ
    \bbeta)^\top (\by - \bZ
    \bbeta)\right\}} {\int \sigma^{-1} \left[\det(\bZ^\top \bZ)\right]^{1/2} \RSS^{1/2} \left(\frac
    {1}{2 \pi \sigma^2}\right)^{n/2} \exp\left\{-\frac{1}{2\sigma^2} (\by - \bZ \bbeta)^\top
    (\by - \bZ
    \bbeta)\right\} d \btheta}.
\end{equation}
Let $p^*$ be the length of $\bbeta$.  The numerator of equation~\eqref{eq:gfd_model} becomes
\begin{align}
    & \int  \sigma^{-1} \left[\det(\bZ^\top \bZ)\right]^{1/2} \RSS^{1/2} \left(\frac
    {1}{2 \pi \sigma^2}\right)^{n/2}\exp\left\{-\frac{1}{2\sigma^2} (\by - \bZ \bbeta)^\top (\by - \bZ
    \bbeta)\right\} q^{p^*} d \btheta  \nonumber \\
    =& (2 \pi)^{(p^*-n)/2} \RSS^{1/2}  \int \sigma^{p^*-n-1} \exp\left(-\frac
    {\RSS} {2\sigma^2}\right) q^{p^*} d\sigma \nonumber \\
    =& (2 \pi)^{(p^*-n)/2}  2^{(n-p^*-2)/2} \RSS^{(p^*-n + 1)/2} \Gamma
    \left(\frac{n-p^*}{2}\right) q^{p^*}.
\end{align}
Thus, the generalized fiducial probability $r(M)$ of any candidate model $M$ is
\begin{align}
\label{eq:model_probability}
r(M) \propto R(M) = (2 \pi)^{(p^*-n)/2}  2^{(n-p^*-2)/2} \RSS^{(p^*-n + 1)/2} \Gamma \left(\frac{n-p^*}{2}\right) q^{p^*}.
\end{align}

%  I am skipping some math here, the fiducial probability of model $M$, denoted by $r(M)$, is proportional to
% \begin{align}
%    (2 \pi)^{(p^*-n)/2}  2^{(n-p^*-2)/2} RSS^{(p^*-n + 1)/2} \Gamma
%     \left(\frac{n-p^*}{2}\right) \times q^{p^*}
% \end{align}
% where $q$ is a turning parametric which roughly tells us the rate of sparsity. For example, $q=0.5$
% encourages roughly 50\% of the predictors to be zero.

 % \color{red}

\subsection{Generating Fiducial Samples}
\label{samples}
This subsection describes how to practically generate fiducial samples $(M, \sigma, \bbeta)$ for the current nonparametric additive modeling problem.

First, to reduce the ``search space'', we consider only candidate models from a subset $\mathcal{M}^*$ of $\mathcal{M}$.  This subset $\mathcal{M}^*$ should contain only candidate models with non-negligible values of $r(M)$.  The way we obtain $\mathcal{M}^*$ is to apply group Lasso \citep{Yuan-Lin06} to the spline representation in~(\ref{eq:spline_rep}), in a manner described below.  Notice that group Lasso is used here as it enforces that all $\beta_{jk}$'s with the same $j$ to be zero or nonzero simultaneously.

Without the loss of generality, we assume that the first $m_0$ functions $f_j$'s in~(\ref{eq:additive_model}) are nonzero. Let $\bbeta_j=(\beta_{j1},\dots,\beta_{jh_n})^\top$ for $j=1\dots,p$, then $\bbeta=(\bbeta_1,\dots,\bbeta_p)^\top$. The group Lasso estimator $\hat{\bbeta}$ is the minimizer of
$$L(\bbeta)=\lVert \bY - \bZ \bbeta \rVert_2^2+\lambda\sum_{j=1}^p \lVert \bbeta_j \rVert_2 $$
subject to the constraint that
$$\sum_{i=1}^n\sum_{k=1}^{h_n}\beta_{ik}\varphi_{k}(Z_{ij})=0, $$
where $\lambda$ is a penalty parameter.  The constraint can be dropped if we initially center the response and the basis functions.  Changing the values of $\lambda$ will lead to a sequence of fitted models; i.e., a solution path.  Those fitted models that are on the solution path of group Lasso are taken as candidate models for $\mathcal{M}^*$.  For the purpose of not missing any candidate models with non-negligible $r(M)$ values, we repeat the group Lasso procedure to a number of bootstrapped data and take all the fitted models that lie on the solution paths as $\mathcal{M}^*$.  In this way the size of $\mathcal{M}^*$ is substantially smaller than the size of $\mathcal{M}$, and we expect $\sum_{M\in \mathcal{M}^*} r(M)$ to be very close to 1.

For each $M\in \mathcal{M}^*$, we can compute
\begin{align*}
R(M)=(2 \pi)^{(m-n)/2}  2^{(n-m-2)/2} \RSS^{(m-n + 1)/2} \Gamma
\left(\frac{n-m}{2}\right) \times q^{m}
\end{align*}
with $m$ as the number of nonzero functions in $M$.  The generalized fiducial probability $r(M)$ can then be well approximated by
\begin{align}
\label{eq:rM}
r(M)\approx\frac{R(M)}{\sum_{M^*\in \mathcal{M}^*}R(M^*)}.
\end{align}

For a given model $M$, $\sigma$ and $\bbeta$ can then be sampled from, respectively,
\begin{align}
\label{eq:RSS_dist}
\mbox{RSS}_M/\sigma^2 \sim \chi^2_{n-m}
\end{align}
and
\begin{align}
\label{eq:beta_dist}
\bbeta \sim N(\hat \bbeta_{\rm ML}, \sigma^2 (\bZ_M^\top \bZ_M)^{-1}),
\end{align}
where $\mbox{RSS}_M$ is the residual sum of squares of the candidate model $M$, $\bZ_M$ is the design matrix of $M$, and $\hat \bbeta_{\rm ML}$ is the MLE of $\bbeta$ for $M$.

To summarize, we can generate a fiducial sample $(\tilde{M}, \tilde{\sigma}, \tilde{\bbeta})$ by first drawing a model $\tilde{M}$ from (\ref{eq:rM}), and then $\tilde{\sigma}$ and $\tilde{\bbeta}$ from (\ref{eq:RSS_dist}) and (\ref{eq:beta_dist}), respectively.  \newstuff{Notice that in the above no computationally intensive technique like MCMC is required so the generation of a fiducial sample is relatively fast.  Using a 2018 MacBook Pro with a data set of $n=400$ and $p=600$, the proposed method typically takes around 50 seconds to generate $10^5$ fiducial samples.}

\newstuff{
Lastly we discuss the practical choice of $K$, the number of knots.  It is widely known $K$ will introduce bias if its value is too small, or it will inflate the variance if it is too large.  Our experience is that, as long as $K$ is larger than a certain value, the resulting estimates are very often similar (and acceptable), as the use of group Lasso will shrink those insignificant knots to zero.  From a theoretical standpoint, the calculations of \cite{Lai-et-al12} suggest that $K$ should be of order $\log(n)$.  So in practice we recommend choosing $K$ as the smallest integer larger than $\log(n)$.  Table~\ref{var_coverage} and~\ref{ey_coverage} below suggest that the numerical results are relatively insensitive to the choice of $K$ (as long as $K$ is large enough).
}

\subsection{Point Estimates, Confidence Intervals and Prediction Intervals}
\label{inference}

Repeating the above procedure multiple times will result in a fiducial sample for $(M, \sigma, \bbeta)$ which can be used for inference, in a manner similar to that for a Bayesian posterior sample.  Instead of selecting one single model, the $r(M)$ approximated in~(\ref{eq:rM}) estimates how likely each candidate model would be the true model; this affects the models being selected in the fiducial sample.  For $\sigma$, one can use the average or median of all $\tilde{\sigma}$'s as a point estimate, and the $\alpha/2$ and $1-\alpha/2$ percentiles to construct a $100(1-\alpha)\%$ confidence interval.  Similarly, a confidence interval for $E(Y_i|\bx_i)$ given the observation $(\bx_i,Y_i)$ can be found by computing the percentiles from $\bz_s \tilde{\bbeta}$, where $\bz_s$ is the spline representation of $\bx_i$.  In addition, prediction intervals for $\bY$ can be obtained by taking the percentiles from $\bZ\tilde{\bbeta}+\tilde{\sigma} \bW$, where $\bW\sim N(\bf{0},I_n)$.

\newstuff{
However, constructing confidence bands for the $f_j$'s is a less trivial task, as it is possible that any particular $f_j$ would appear only in some but not all of the fiducial samples.  To handle this issue, we use the following strategy.  First from the fiducial samples, we identify the model $\hat{M}$ that appears most.  If $\hat{M}$ appears for more than 50\% of the times (which has always been the case for all the simulated data sets and the real data example examined), we treat it as the selected model and declare all its non-zero $\hat{f}_j$'s as significant.  For each of these selected function $f_j$'s, we then form a confidence band by finding the corresponding percentiles from $\bZ_f\bbeta_f$ where $\bZ_f$ and $\bbeta_f$ are, respectively, the spline representation and the part of $\tilde{\bbeta}$ corresponding to this selected function.
}

\section{Theoretical Properties}
\label{sec:theory}

This section presents some asymptotic properties of the above generalized fiducial based method.  We assume that $p$ is diverging and the theoretical properties are established under the following conditions.

(A1) Let $\mathcal{H}$ be the class of functions $h$ on $[a,b]$ which satisfies a Lipschitz condition of order $\alpha$:
$$
|h^{(k)}(s)-h^{(k)}(t)|\leq C|s-t|^{\alpha}\,\,\,\text{for}\,s,t\in[a,b],$$
where $k$ is a nonnegative integer and $\alpha \in (0,1]$ so that $d=k+\alpha>0.5$. Then $f_{j} \in \mathcal{H}$ for $1\leq j\leq q$.

(A2) Let $a=\xi_0<\xi_1< \xi_2<\dots<\xi_K<\xi_{K+1}=b$ denote a partition of $[a,b]$ into $K+1$ subintervals where the $t$-th subinterval $I_t=[\xi_{t-1},\xi_{t})$ for $t=1,\dots,K$ and $I_{K+1}=[\xi_{K},\xi_{K+1}]$. We assume that these knots are not overly sparse; i.e., let $0<\nu<0.5$ and $K=n^{\nu}$ be a positive integer such that $\max_{1\leq t\leq K+1}|\xi_{t}-\xi_{t-1}|=O(n^{-\nu})$.

(A3) There exists a constant $c_0$ such that $\min_{1\leq j\leq q}\lVert f_j\rVert_2\geq c_0$, where $\lVert f\rVert_2=[\int_a^b f^2(x)dx]^{1/2}$ whenever the integral exists.

 (A4) $\bX$ has a continuous density and there exist constants $C_1$ and $C_2$ such that the density function $g_j$ of $X_j$ satisfies $0<C_1\leq g_j(X)\leq C_2>\infty$.

 (A5) Let $m$ and $m_0$ be the number of nonzero functions selected for models $M$ and $M_0$, respectively.  Then $p^*=h_nm$ for model $M$. We consider only $M\in\mathcal{M}$ where $\mathcal{M}=\{M: m\leq km_0\}$ for a finite constant $k>1$; i.e., the model whose size is comparable to the true model.

 (A6) Let $\Delta (M)=\|\bmu-\bH_M\bmu\|$ where $\bmu=\bZ_{M_0}\bbeta_{M_0}$. We assume the following identifiability condition:
 $$ \lim_{n\rightarrow \infty}\min \left\{\frac{\Delta(M)}{h_nm_o\log p}: M_0\notin {\mathcal M}, m\leq km_0 \right\}=\infty $$
 This condition ensures that the true model can be differentiated from the other models.

 (A7) There exists a variable screening procedure to reduce the size of $\mathcal{M}$ when $p$ is too large in practice.  Denote the class of candidate models resulting from the screening procedure by $\mathcal{M}^*$.  Existence follows from
 \begin{align}
 \label{eq:a7}
 P(M_0\in \mathcal{M}^*)\rightarrow 1 \quad \mbox{and} \quad \log(|\mathcal{M}_j^*|)=o(h_nj\log n),
 \end{align}
 where $\mathcal{M}_j^*$ denotes the set of all sub-models in $\mathcal{M}^*$ of size $j$. These two limiting criteria ensure that the true model is contained in $\mathcal{M}^*$ and the size of the model space $\mathcal{M}^*$ is not too large.

 The following theorem summarizes our main results and its proof can be found in the appendix.
 {\thm{
\label{thm:1}
        Assume A1-A6 hold.  As $n\rightarrow \infty$, $p\rightarrow \infty$, $h_nm_0\log (p)=o(n)$, $\log (h_nm_0)/\log (p)\rightarrow \delta$ and $-\log(q)/\log(p)=\gamma$, we have
        \begin{align}
        \label{eq:thm1}
        \max_{M\neq M_0, M\in\mathcal{M}}\frac{r(M)}{r(M_0)}\xrightarrow{P}0,
        \end{align}
for $1+\delta<\gamma<C$ with $C$ being a constant.}

        Moreover if A7 also holds, with the same $\gamma$ we have
        \begin{align}
        \label{eq:thm2}
        r(M_0)\xrightarrow{P} 1
        \end{align}
        over the class $\mathcal{M}^*$.
}

Theorem~\ref{thm:1} states that the true model $M_0$ has the highest generalized fiducial probability amongst all the candidate models under some regularity conditions, and if in addition equation~(\ref{eq:a7}) holds, the true model will be selected with probability tending to $1$. Note that equation~({\ref{eq:thm1}}) does not imply~({\ref{eq:thm2}}) in general since we assume a diverging $p$. Here $\gamma$ plays a role similar to that of the tuning parameter in EBIC of \cite{Luo-Chen13}, which controls a penalty for the size of the class of submodels and it must fall within a specified range to ensure that the generalized fiducial distribution is consistent.

In practice, we use group Lasso on bootstrapped data to generate candidate models as discussed in Section~\ref{samples} above.  The resulting model space satisfies equation~(\ref{eq:a7}), since group Lasso is selection consistent for some $\lambda$ as shown in \cite{nardi2008}.  Theorem~\ref{thm:1} also implies that statistical inference based on the generalized fiducial density~(\ref{eq:generalized_fiducial_density}) will retain the exact asymptotic frequentist property as shown in Theorem~$2$ and Theorem~$3$ of \cite{Hannig-et-al16}, which ensure the consistency of our inferential procedure.

\newstuff{
We close this section with the following two remarks.  First, in general, nonparametric function estimators are biased.  We handle this issue by imposing some restrictions on the true functions $f_j$'s.  Lemma 1 implies that these functions can be well approximated by polynomial splines adopted in the proposed method (i.e., the bias vanishes asymptotically).  With these somewhat strong restrictions, we are able to obtain the above theoretical results.  The second remark is that establishing theoretical results for the estimation of $\beta_{jk}$'s is more challenging.  One reason is that the optimal number of $\beta_{jk}$'s can vary for different $f_j$'s and we do not have any method or consistency results for estimating this number under the current setting.  Another difficulty is that, in practice, some fitted models in the fiducial samples contain a certain set of $\beta_{jk}$'s while some other fitted models do not.  This makes it very difficult to even construct confidence intervals for such $\beta_{jk}$'s, let alone conduct rigorous study of any theoretical property.
}

\section{Empirical Properties}
\label{sec:numerical}
This subsection investigates the empirical properties of the proposed method via numerical experiments and a real data example.

\subsection{Simulation Experiments}
Following the simulation settings in \cite{Huang-et-al10}, we use the model
\[
y_i = \sum_{j=1}^{p} f_j(x_{ij}) + \varepsilon_i, \quad i=1, \ldots n, \quad \varepsilon_i \sim \mbox{i.i.d.} \, N(0, \sigma^2)
\]
to generate simulated data, where
\begin{eqnarray*}
f_1(x) & = & 5 x,\\
f_2(x) & = & 3 (2x-1)^2, \\
f_3(x) & = & 4 \sin(2\pi x) /\{2- \sin(2\pi x)\},  \\
f_4(x) & = & 6 \{0.1 \sin(2 \pi x) + 0.2 \cos (2 \pi x)\} + 0.3\sin^2(2\pi x) + 0.4\cos^3(2\pi x) + 0.5\sin^3(2\pi x), \\
f_j(x) & = & 0 \quad \mbox{for} \quad 5\le j \le p,
\end{eqnarray*}
and the noise variance $\sigma^2$ is chosen such that the signal-to-noise ratio is greater than $1$ for each nonzero function.

%Specifically, $f_1(x) = 5t, f_2(x)=3 (2t-1 )^2, f_3(x)=3 (2t-1 )^2, f_4(x)=6 (0.1 \sin(2 \pi t) + 0.2 \cos (2 \pi t)) + 0.3\sin^2(2\pi t) + 0.4\cos^3(2 \pi t) + 0.5\sin^3(2 \pi t)$ and $f_j(x)=0$ for $5\le j \le p$.

%So we are going to draw some samples of $(M,\sigma^2,\beta)$, then construct the confidence
%bands. In the simulations, $n=200$ and $p=1000$.

For each set of simulated data, we first use B-spline expansions to transform our data to representation (\ref{eq:spline_rep}). Then a set of candidate models $\mathcal{M}$ are generated by using group Lasso on the transformed data and $10$ sets of bootstrapped data. For each $M$, we run a simple linear regression to obtain $\RSS_M$ and compute the fiducial probability $r(M)$ as shown in (\ref{eq:model_probability}). Then we can draw samples of $(M,\sigma^2,\bbeta)$ based on $r(M)$, (\ref{eq:RSS_dist}) and (\ref{eq:beta_dist}) and construct confidence intervals or bands.

\begin{figure}[ht!]
\includegraphics[scale=0.5]{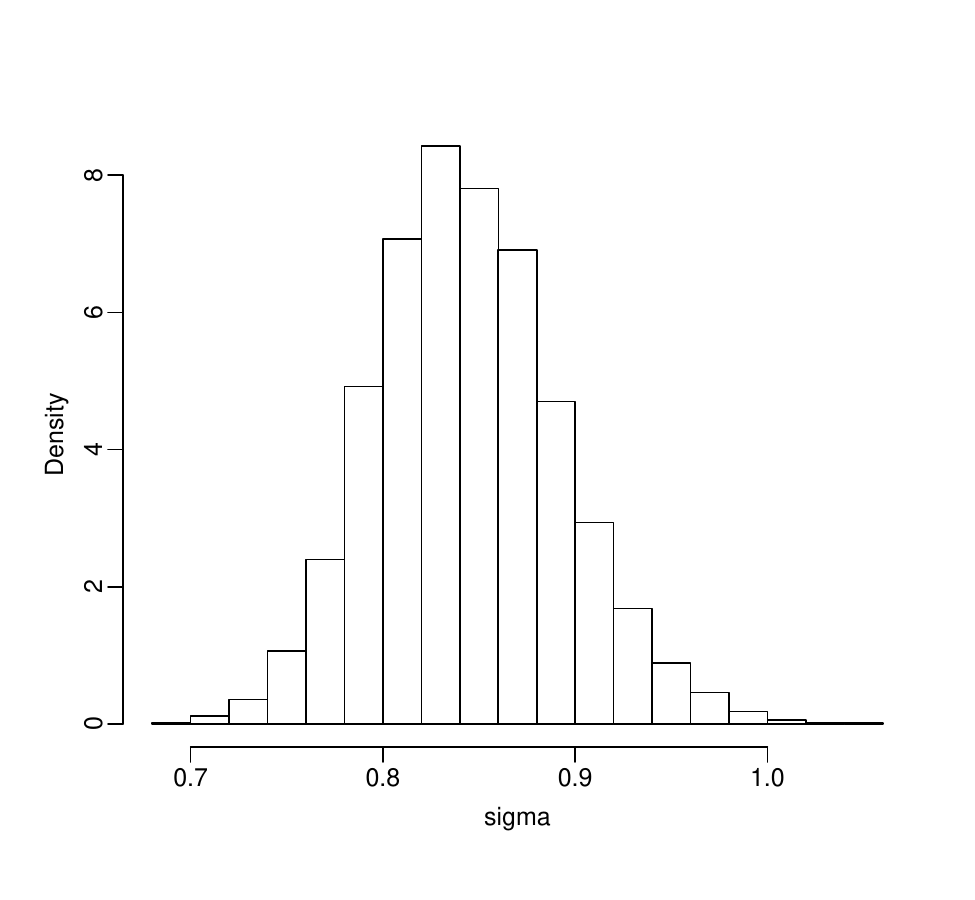}
\hspace*{-1cm}
\includegraphics[scale=0.5]{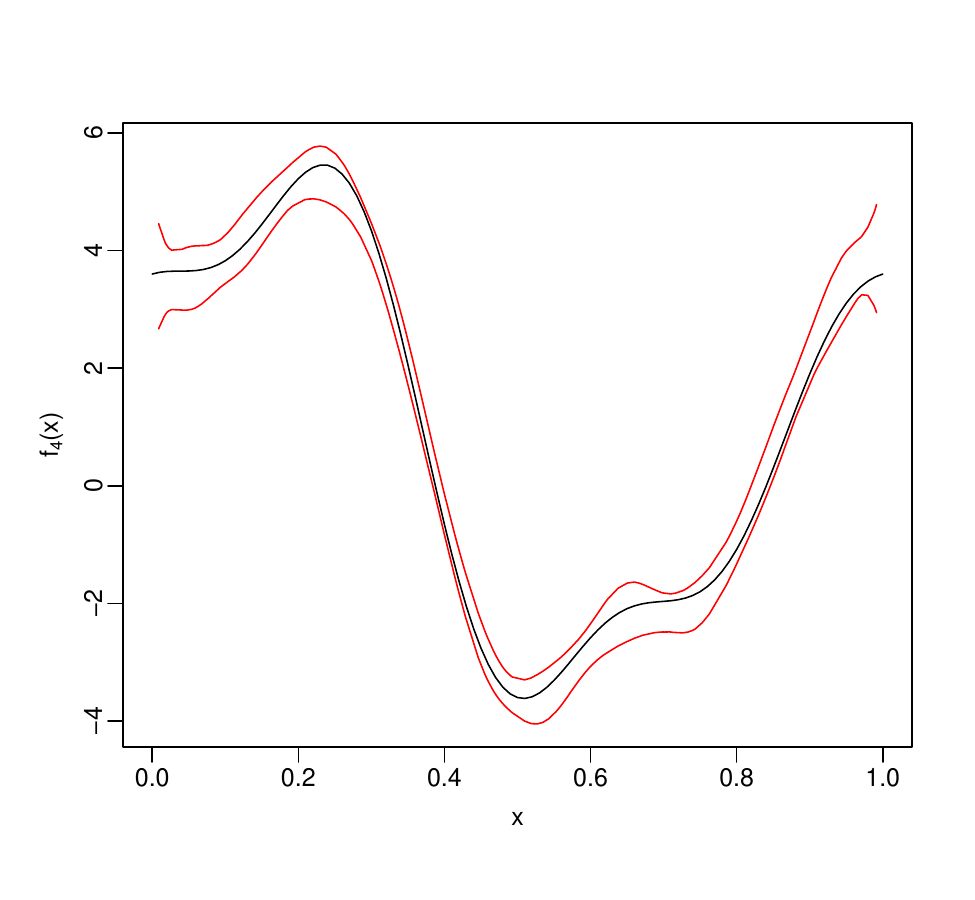}
%\vspace*{-1.2cm}
\caption{Left: histogram of the fiducial samples of $\sigma$.  Right: a 95\% pointwise confidence band of $f_4$.  The black line is the true function while the red lines show the band.}
\label{fig:one.sim}
\end{figure}

Figure~\ref{fig:one.sim} summarizes some results of applying the proposed method to a typical simulated data set with $n=200$, $p=1,000$ and $\sigma=0.8$.  For the B-spline expansion we use $l=3$ and $K=8$, and $10,000$ samples of $(M, \sigma^2, \bbeta)$ are generated.  Using these samples a $95\%$ confidence interval for $\sigma$ is obtained, which is $(0.756, 0.947)$ and includes the true value 0.8.  The left panel in Figure~\ref{fig:one.sim} depicts the histogram of the $10,000$ samples of $\sigma$ which can be seen to be approximately normally distributed.  The right panel shows the $95\%$ pointwise confidence band of $f_4(x)$, where the black line is the true function and the red lines are the two bounds.  We use $f_4(x)$ here since it is the most complicated of the four non-zero functions.  We can see that the confidence band covers the true function very well.

To test the coverage of these confidence intervals, we generate $1,000$ simulated data sets and apply the proposed method to compute the confidence intervals for $\sigma^2$ and the mean function $E(Y_i|\bx_i)$ evaluated at $n$ design points $\bx_i$'s. We compare the performance of our method with the ``oracle'' method which uses the true model and classical theories in linear models based on the spline representation to derive confidence intervals. Different combinations of $n$, $p$, $\sigma$, $l$, $K$ and $\alpha$ are tested and the numerical results are summarized in Table~\ref{var_coverage} and~\ref{ey_coverage}.  The empirical coverage rates are reported together with the average widths of the intervals shown in parentheses.

\begin{table}[ht!]
{\small
%\vspace*{-0.3cm}
        \begin{center}
                \begin{tabular}{ |l|l|l|lll| } \hline
                        &  &  & $90\%$ & $95\%$ & $99\%$ \\ \hline
                        \multirow{6}{*}{$(n,p,\sigma)=(200,1000,1)$} &
                        {$l=3, K=6$} & proposed & $86.40\%$ ($0.392$) & $92.90\%$ ($0.466$) & $95.10\%$ ($0.630$) \\
                        & & oracle & $89.70\%$ ($0.374$) & $95.60\%$ ($0.447$)  & $98.50\%$ ($0.595$)  \\
                        &{$l=3, K=8$}  & & $86.60\%$ ($0.410$)  & $91.20\%$ ($0.501$) & $94.82\%$ ($0.672$) \\
%                        &{$l=3, K=8$}  & & $79.30\%$ ($0.501$)  & $79.80\%$ ($0.620$) & $86.70\%$ ($0.869$) \\
                        &  & & $90.40\%$ ($0.378$) & $94.30\%$ ($0.454$) & $99.10\%$ ($0.609$)\\
                        & {$l=4, K=6$} &  & $86.80\%$ ($0.429$) &  $89.60\%$ ($0.535$) & $94.50\%$ ($0.714$) \\
                        &  &  & $91.60\%$ ($0.376$) & $94.40\%$ ($0.451$) & $98.30\%$ ($0.599$)\\ \hline
                        \multirow{6}{*}{$(n,p,\sigma)=(200,1000,0.8)$} & {$l=3, K=6$}& proposed  & $89.80\%$ ($0.242$)  &$94.40\%$ ($0.286$) & $98.60\%$ ($0.384$)\\
                        &  & oracle & $90.00\%$ ($0.243$) & $94.60\%$ ($0.288$) & $99.10\%$ ($0.384$)\\
                        & {$l=3, K=8$} &  & $88.89\%$ ($0.244$) & $93.00\%$ ($0.295$) & $98.20\%$ ($0.395$) \\
                        &  &  & $89.59\%$ ($0.242$) & $93.50\%$ ($0.292$) & $99.20\%$ ($0.389$)\\
                        & {$l=4, K=6$} &  & $90.40\%$ ($0.241$) & $93.70\%$ ($0.289$)  & $99.10\%$ ($0.383$)\\
                        & &  & $89.80\%$ ($0.242$) & $93.00\%$ ($0.290$) & $99.10\%$ ($0.386$) \\ \hline
                        \multirow{6}{*}{$(n,p,\sigma)=(250,1500,0.8)$} &{$l=3, K=6$} & proposed & $89.50\%$ ($0.207$)  & $94.80\%$ ($0.248$)  & $98.30\%$ ($0.329$) \\
                        &  & oracle & $89.00\%$ ($0.208$) & $94.60\%$ ($0.207$)  & $98.10\%$ ($0.331$) \\
                        & {$l=3, K=8$} &  & $90.90\%$ ($0.210$)  & $94.29\%$ ($0.252$) &$98.80\%$ ($0.333$) \\
                        &  &  & $91.00\%$ ($0.210$) & $94.09\%$ ($0.253$) & $98.60\%$ ($0.334$)\\
                        & {$l=4, K=6$} &  & $88.70\%$ ($0.212$) & $92.69\%$ ($0.253$) & $98.68\%$ ($0.338$)\\
                        & &  & $88.10\%$ ($0.214$) & $92.89\%$ ($0.255$)  & $98.38\%$ ($0.340$) \\ \hline
                \end{tabular}
        \end{center}
}
\caption{Empirical coverage rates of confidence intervals for $\sigma^2$. Numbers in parentheses are average widths of the confidence intervals.}
\label{var_coverage}
%        \vspace*{-0.3cm}
\end{table}

 \begin{table}[ht!]
%        \vspace*{-0.3cm}
{\small
        \begin{center}
                \begin{tabular}{ |l|l|l|lll| } \hline
                        &  &  & $90\%$ & $95\%$ & $99\%$ \\ \hline
                        \multirow{6}{*}{$(n,p,\sigma)=(200,1000,1)$} &
                        {$l=3, K=6$} & proposed & $87.35\%$ ($1.401$) & $93.19\%$ ($1.668$) & $98.05\%$ ($2.205$)\\
                        & & oracle & $88.75\%$ ($1.385$) & $93.92\%$ ($1.648$)  & $98.53\%$ ($2.167$)\\
                        &{$l=3, K=8$}  &  & $86.46\%$ ($1.601$) & $91.39\%$ ($1.923$) & $97.05\%$ ($2.562$) \\
                        &  & & $89.41\%$ ($1.524$) & $94.38\%$ ($1.817$) & $98.76\%$ ($2.396$) \\
                        & {$l=4, K=6$} &  & $87.70\%$ ($1.489$) & $93.36\%$ ($1.789$)  & $98.17\%$ ($2.353$)  \\
                        &  &  & $89.08\%$ ($1.452$) & $94.40\%$ ($1.731$) & $98.77\%$ ($2.272$)\\ \hline
                        \multirow{6}{*}{$(n,p,\sigma)=(200,1000,0.8)$} &{$l=3, K=6$} & proposed & $89.08\%$ ($1.170$)  &$93.63\%$ ($1.328$) & $98.50\%$ ($1.757$)\\
                        &  & oracle & $89.00\%$ ($1.167$) & $93.55\%$ ($1.323$) & $98.57\%$ ($1.741$)\\
                        & {$l=3, K=8$} &  & $89.14\%$ ($1.225$) & $94.38\%$ ($1.467$) & $98.72\%$ ($1.937$) \\
                        &  &  & $89.31\%$ ($1.220$) & $94.45\%$ ($1.457$) & $98.81\%$ ($1.916$)\\
                        & {$l=4, K=6$} &  & $88.86\%$ ($1.168$) & $94.13\%$ ($1.395$)  & $98.72\%$ ($1.839$) \\
                        & &  & $88.83\%$ ($1.165$)  & $94.10\%$ ($1.389$) &  $98.66\%$ ($1.825$)\\ \hline
                        \multirow{6}{*}{$(n,p,\sigma)=(250,1500,0.8)$} &{$l=3, K=6$} & proposed &$88.17\%$ ($0.991$)  & $93.62\%$ ($1.183$)  & $98.47\%$ ($1.557$)\\
                        &  & oracle & $88.14\%$ ($0.989$)  & $93.53\%$ ($1.179$)  & $98.46\%$ ($0.983$)\\
                        & {$l=3, K=8$} &  & $89.33\%$ ($1.092$) & $94.38\%$ ($1.306$) & $98.79\%$ ($1.716$)\\
                        &  &  & $89.28\%$ ($1.090$) & $94.32\%$ ($1.302$) & $98.76\%$ ($1.707$)\\
                        & {$l=4, K=6$} &  & $87.48\%$ ($1.050$) & $93.04\%$ ($1.251$) &$98.33\%$ ($1.651$) \\
                        & &  & $87.41\%$ ($1.048$) & $92.98\%$ ($1.247$) &  $98.29\%$ ($1.643$)\\ \hline
                \end{tabular}
        \end{center}
}
\caption{Empirical coverage rates of confidence intervals for E$(Y_i|\brm{x}_i)$. Numbers in parentheses are the average widths of the confidence intervals.}
\label{ey_coverage}
%        \vspace*{-0.3cm}
 \end{table}

\begin{figure}[ht!]
\includegraphics[scale=0.75]{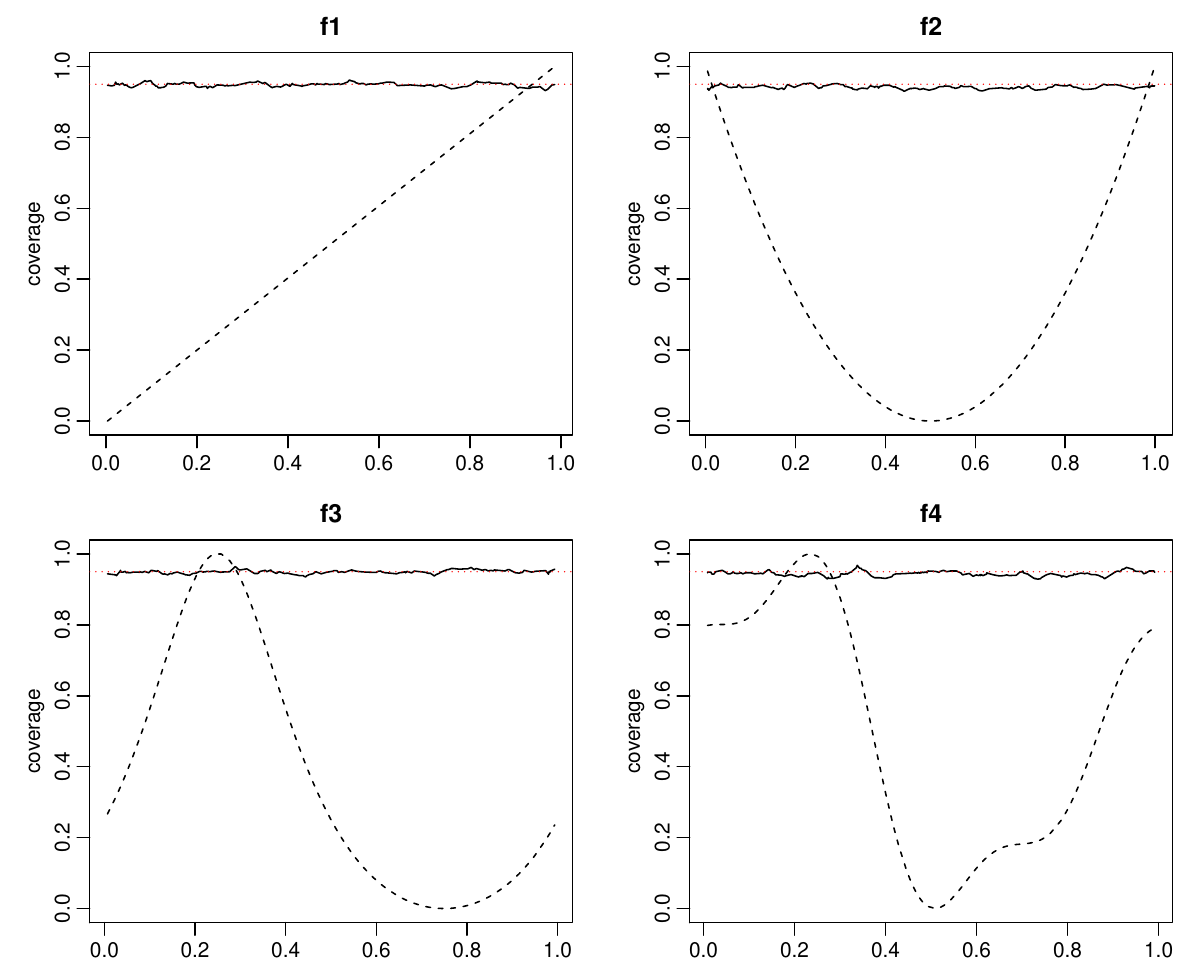}
\caption{Empirical coverage rates for each non-zero function with experimental parameters $n=200$, $p=1,000$, $\sigma=0.8$, $\alpha=5\%$, $l=3$ and $K=8$.}
\label{fig:coverage}
\end{figure}

To evaluate the performance visually, we also plot the empirical coverage rates of all four non-zero functions for one combination of experimental parameters; see Figure~\ref{fig:coverage}.  In each panel the black dashed line depicts the true value of the function, the horizontal red dashed line is the target confidence level ($95\%$ in this case) while the black solid line represents the empirical coverage rates.  One can see that these rates are very close to the target confidence level.

\clearpage

\subsection{Comparison with an existing method}
\newstuff{
This subsection compares the performances of the proposed method with those from the kernel-sieve hybrid estimator developed by \citet{Lu-et-al15}.  As with the settings in \citet{Lu-et-al15}, the same set of test functions in the previous subsection are used with $\sigma^2=1.5^2$, $p=600$, and $n\in\{400,500,600\}$.  For the kernel-sieve hybrid estimator, we follow the parameter selection used in \citet{Lu-et-al15}.  For the proposed method, we set $l=3$ and $K=8$.
}

\begin{figure}[ht]
    \centering
    \includegraphics[width=0.325\textwidth]{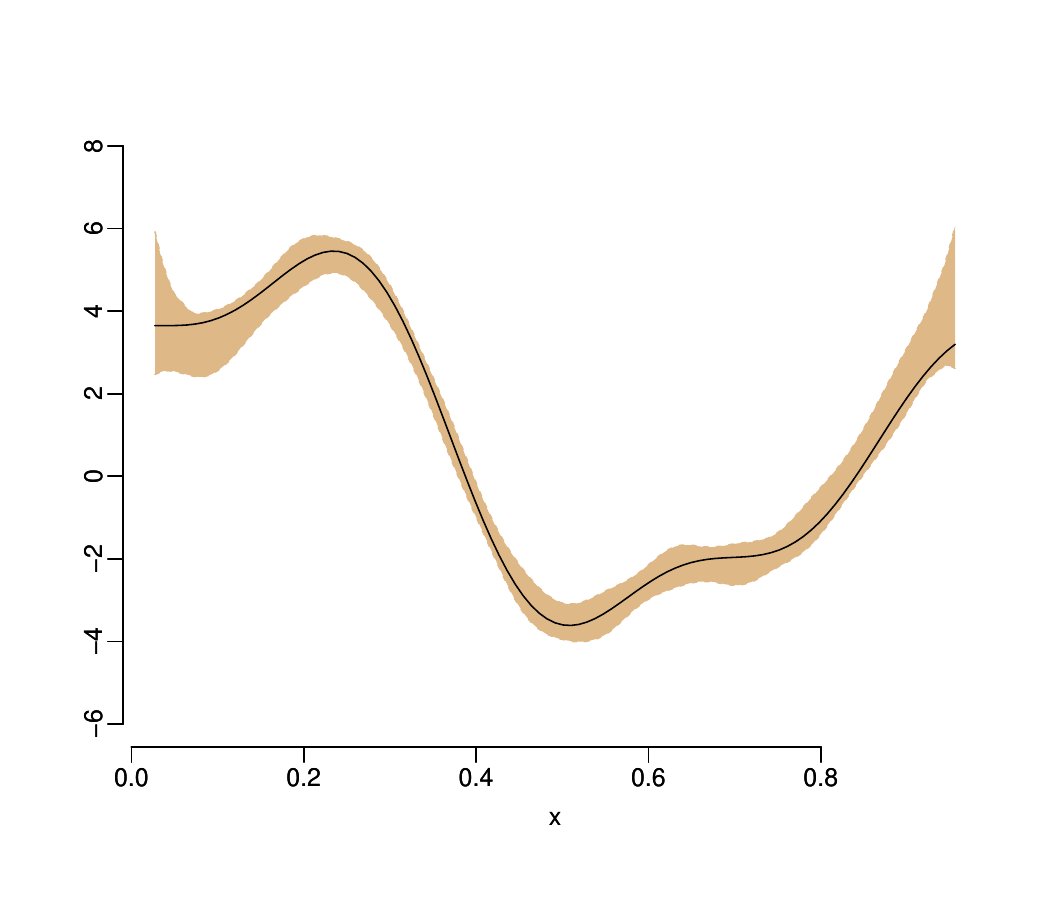}
    \includegraphics[width=0.325\textwidth]{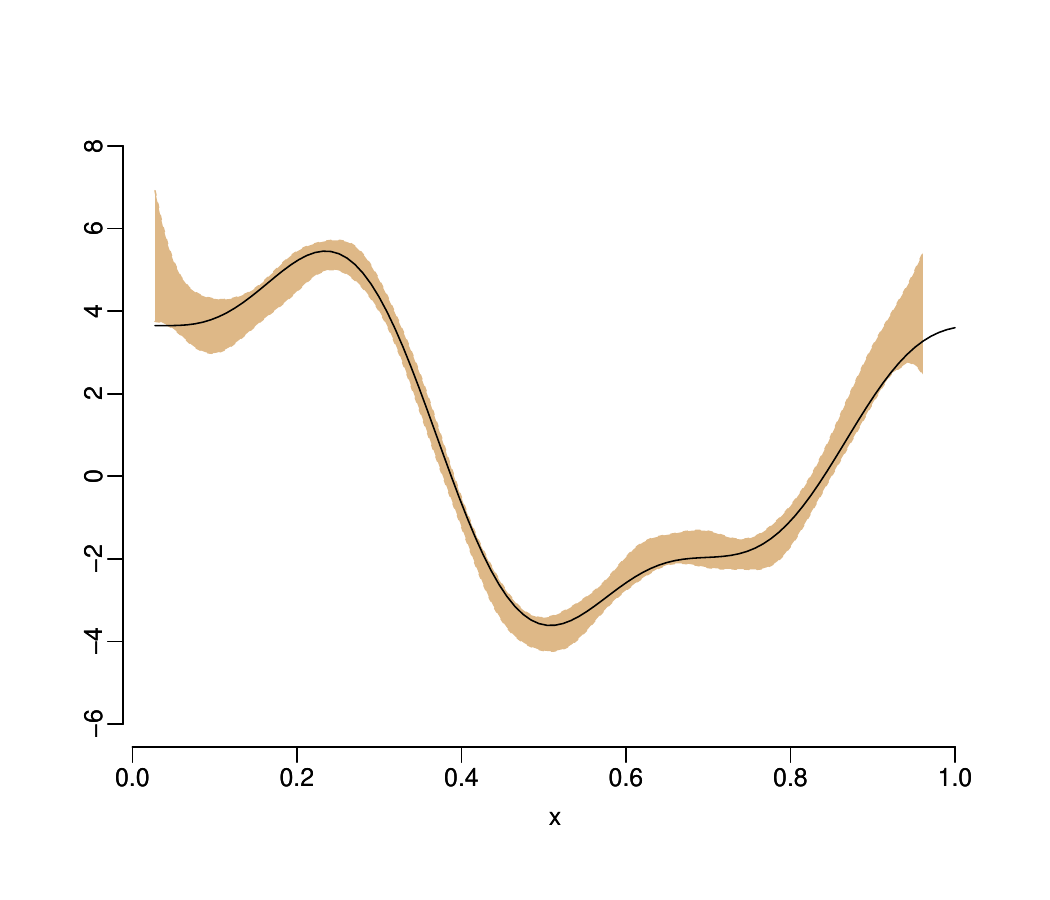}
    \includegraphics[width=0.325\textwidth]{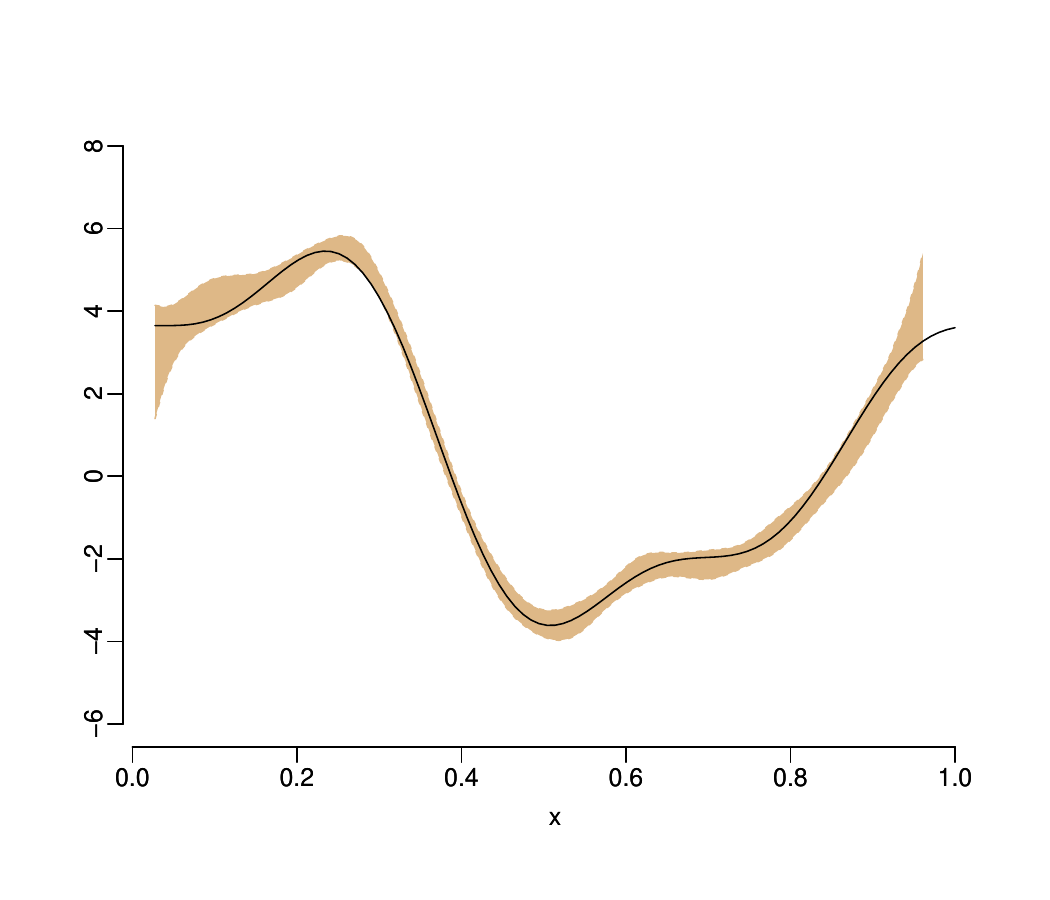}
\newstuff{
  \caption{Plots of $95\%$ confidence bands produced by the proposed method.  These bands are targeting $f_4(x)$. From left to right, $n=400, 500, 600$, respectively.}
  }
    \label{fig:gfi}
\end{figure}

\begin{figure}[ht]
    \centering
    \includegraphics[width=\textwidth]{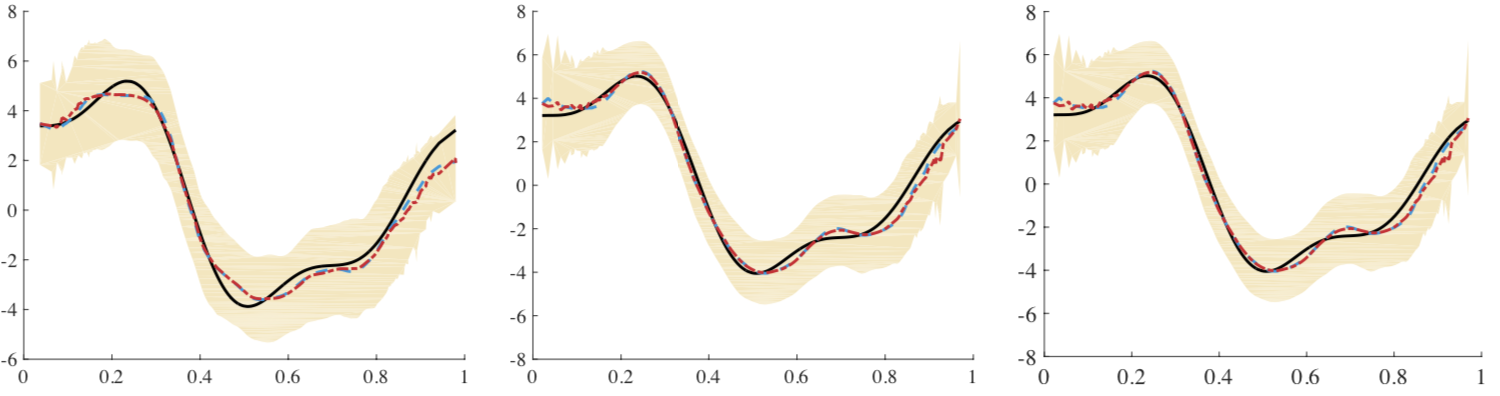}
\newstuff{
  \caption{Similar to Figure~\protect\ref{fig:gfi} but for the kernel-sieve hybrid estimator.}
  }
    \label{fig:ks}
\end{figure}

\newstuff{
The empirical coverage rates that the 95\% confidence bands cover the true $f_4(x)$ on the first 100 data points are computed based on 500 repetitions.  These coverage rates are summarized in Table~\ref{tab:cov}.  One can see that both methods produce very reasonable results, with those from the proposed method being closer to the nominal significance level.  For visual evaluation, 95\% confidence bands for typical data sets are displayed in Figure~\ref{fig:ks} and Figure~\ref{fig:gfi}.  These plots suggest that the proposed method produces tighter confidence bands.
}
\begin{table}[ht]
\centering
\label{tab:cov}
\newstuff{
\begin{tabular}{c|c|c} \hline
$n$ & Generalized Fiducial Inference & Kernel-sieve Hybrid Estimator\\ \hline
400 & 94.9 &  91.6  \\ \hline
500 & 94.8 &  93.2  \\ \hline
600 & 95.3 &  93.6  \\ \hline
\end{tabular}
}
\newstuff{
  \caption{Empirical coverage rates of 95\% confidence bands targeted for $f_4(x)$.}
  }
\end{table}

\newstuff{
Lastly we report the execution times for the methods.  For a typical data set of $n=400$, the proposed method and the kernel-sieve hybrid estimator take about 52s and 903s, respectively, to finish.  The code for the proposed method is written in $R$ while the code for the hybrid estimator (kindly provided by one of its authors) is in Matlab.  The machine is a 2018 MacBook Pro with a 2.3 GHz Intel Core i5 processor.
}

\subsection{Real data example}
This subsection presents a real data analysis on the riboflavin (vitamin B$_2$) production data set which is available in Supplementary Section~A.1 of \cite{buhlmann-et-al2014}. The response variable is the logarithm of the riboflavin production rate in Bacillus subtilis for $n=71$ samples while there are $p=4,088$ covariates measuring the logarithm of the expression level of $4,088$ genes. \cite{buhlmann-et-al2014} and \cite{javanmard2014} use linear models to detect significant genes that potentially affect riboflavin production.  \cite{buhlmann-et-al2014} locate the gene {\fontfamily{cmss}\selectfont YXLD-at} while \cite{javanmard2014} identify the two genes {\fontfamily{cmss}\selectfont YXLD-at} and {\fontfamily{cmss}\selectfont YXLE-at} as significant. Here, instead of using a simple linear model, we assume a nonparametric additive model and apply the GFI methodology to select significant genes.

Following \cite{buhlmann-et-al2014}, we first adopt a screening procedure and use only the $100$ genes with the largest empirical variances.
%Then we use our algorithm to generate candidate models based on $1000$ bootstrapped data and compute the corresponding fiducial densities. Here we use $K=2,l=3$ and $q=0.02$.
We then apply the proposed method with $K=2$ and $l=3$ to the screened data set and obtain 10,000 fiducial samples for $(M, \sigma, \bbeta)$.  It turns out with $63.2\%$ fiducial probability, {\fontfamily{cmss}\selectfont YXLD-at} and {\fontfamily{cmss}\selectfont YBFG-at} are jointly selected while with $28.4\%$ fiducial probability, {\fontfamily{cmss}\selectfont YXLD-at} and {\fontfamily{cmss}\selectfont XHLA-at} are jointly selected.  In other words the proposed method is capable of detecting {\fontfamily{cmss}\selectfont YXLD-at} which is considered significant in most previous analyses of this data set.  Also, with the 10,000 fiducial samples we construct a $95\%$ confidence interval for $\sigma$, which is $(0.43, 0.62)$.

%Following \cite{buhlmann-et-al2014}, we adopted a screening procedure use the $100$ genes with the largest empirical variances for ease of reproduction. Then we use our algorithm to generate candidate models based on $1000$ bootstrapped data and compute the corresponding fiducial densities. Here we use $K=2,l=3$ and $q=0.02$. It turns out the with $63.2\%$ probability, {\fontfamily{cmss}\selectfont YXLD-at} and {\fontfamily{cmss}\selectfont YBFG-at} are selected while with $28.4\%$ probability, {\fontfamily{cmss}\selectfont YXLD-at} and {\fontfamily{cmss}\selectfont XHLA-at} are selected. Here we are able to find {\fontfamily{cmss}\selectfont YXLD-at} which is considered significant in most previous analysis on this data set. Then we draw $10,000$ fiducial samples of $\sigma$ and $\brm{\beta}$ according to the candidate models randomly sampled using the fiducial densities and the resulting $95\%$ confidence interval for $\sigma$ is $(0.43, 0.62)$.

\begin{figure}[ht!]
        \centering
\includegraphics[scale=0.55]{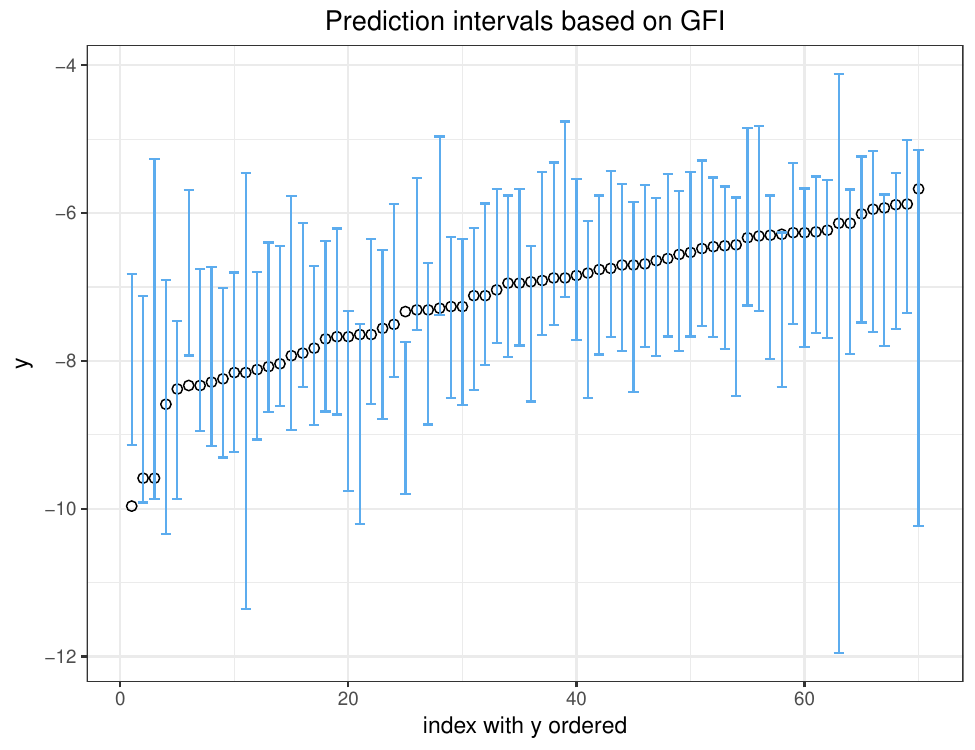}
%        \vspace*{-0.2cm}
        \caption{95\% prediction intervals, denoted as blue error bars, for the responses $Y_i$'s, denoted as black circles.  For clarity the $Y_i$'s are sorted in ascending order.}
        \label{fig:error_bar}
\end{figure}

\begin{figure}[ht!]
        \centering
\includegraphics[scale=0.55]{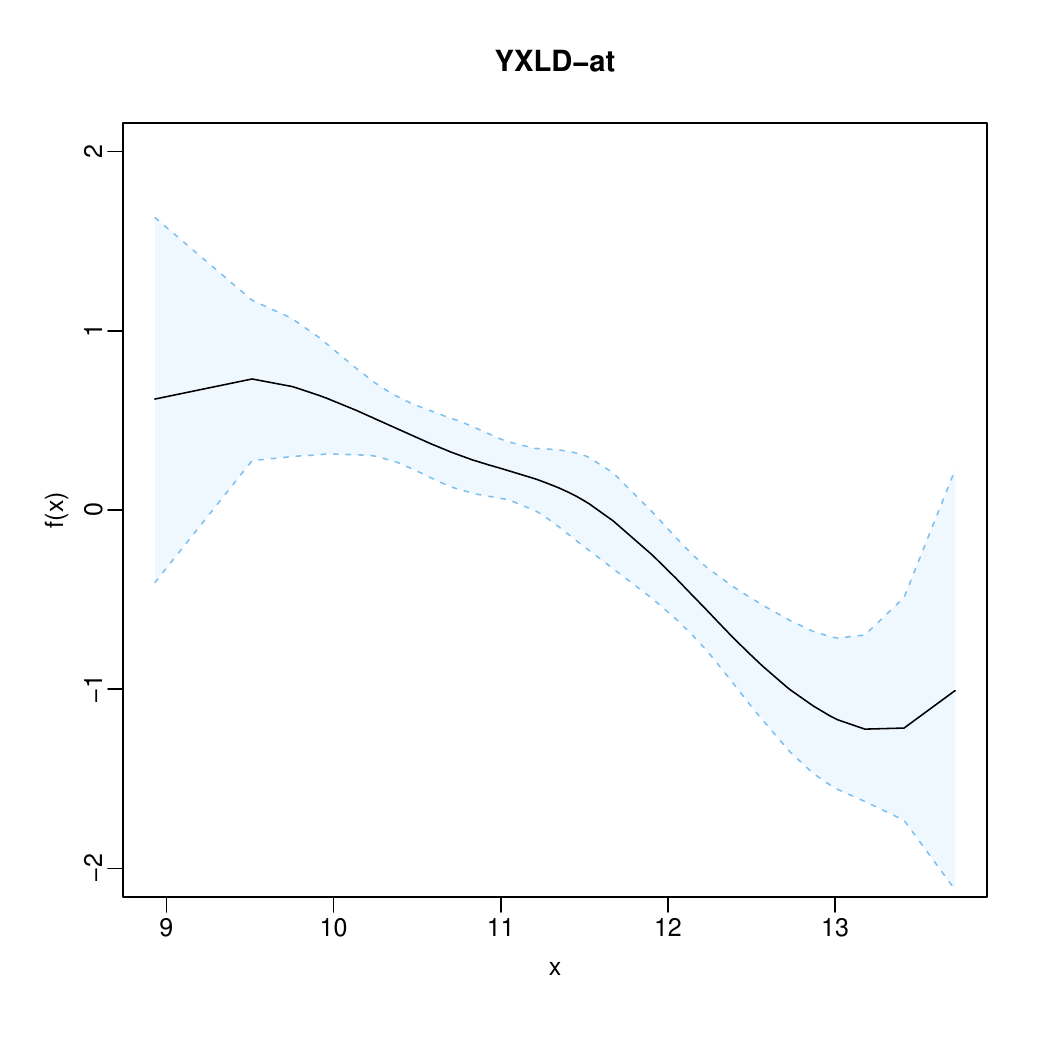}
%        \vspace*{-1.2cm}
        \caption{A 95\% pointwise confidence band for YXLD\_at.  The black solid line is the median of the fiducial samples and the dashed blue lines represent the confidence band.}
        \label{fig:conf_bands}
\end{figure}

From the fiducial samples of $(M, \sigma, \brm{\beta})$, we also compute the leave-one-out 95\% prediction intervals for the responses $Y_i$'s and the results are displayed in Figure~\ref{fig:error_bar}.  Note that for clarity the $Y_i$'s are sorted in ascending order.  From the plot we can see that $68$ out of $71$ prediction intervals cover the value of $Y_i$'s, which is around $95.8\%$.  We also compute the $95\%$ pointwise confidence band for {\fontfamily{cmss}\selectfont YXLD-at} which is shown in Figure~\ref{fig:conf_bands}.  For the $i$th function, such a confidence band can be constructed by using the quantiles from $\bZ_i\brm{\beta}_i$ where $\bZ_i$ and $\brm{\beta}_i$ are, respectively, the design matrix and coefficients corresponding to the $i$th function after the B-spline expansion. In Figure~\ref{fig:conf_bands} the black solid line is the median among all the samples as the true function is not available for real data, while the dashed lines represent the confidence band.  This plot strongly suggests that this gene is indeed significant and the overall trend is more complicated than a simple straight line.  We note that, although many previous methods based on high dimensional linear regression have successfully identified this gene as significant, these methods fail to provide any flexible estimate for the trend, such as the one in Figure~\ref{fig:conf_bands}.

\section{Conclusion}
\label{sec:conclude}
In this paper we adopted a generalized fiducial inference methodology to perform statistical inference on sparse high dimensional nonparametric additive models.  In particular we developed a procedure to generate fiducial samples based on the generalized fiducial distribution of a set of candidate models obtained from group Lasso, and to construct various confidence intervals and prediction intervals by making use of these samples. The developed inferential procedure was shown to have an exact asymptotic frequentist property under some regularity conditions, which was confirmed by its promising performance in numerical simulations.
%Our work considers a spline representation of the nonparametric additive models instead of the original form and the approximation error is not included in our discussion. In future work, we could take the error brought by spline approximation into account and further refine the theoretical results.
We note that the current framework can in principle be extended to other more complicated and flexible models in high dimensional settings, such as the generalized nonparametric additive models.

\subsection*{Acknowledgment}
The authors are most grateful to the reviewers, the associate editor and the editor for their most constructive and helpful comments which led to a much improved version of the paper.

\appendix

\section{Technical Details}
This appendix provides technical details, including the proof for Theorem~\ref{thm:1}.  We begin with three lemmas.

\subsection{Lemmas}
{ \lemma \label{spline}

Let $\mathcal{F}$ be the class of functions $f$ on $[a,b]$ which satisfies:
$$
|f^{(k)}(s)-f^{(k)}(t)|\leq C|s-t|^{\alpha}\,\,\,\text{for}\,s,t\in[a,b],
$$
where $k$ is a nonnegative integer and $\alpha \in (0,1]$ so that $d=k+\alpha>0.5$.

Let $S^0_n$ denote the space of centered polynomial splines. Suppose that $f\in\mathcal{F}$, E$f(Z_j)=0$ and $h_n=O(n^{1/(2d+1)})$, then there exists $f_n\in S^0_n$ satisfying
$$
\|f_n-f\|_2=O_p(h_n^{-d})=O_p(n^{-d/(2d+1)}).
$$

This lemma is proved in \cite{Huang-et-al10} and it indicates that the $f_j$'s can be well approximated by polynomial splines under certain smoothness assumptions. Therefore, the representation we consider in equation~\eqref{eq:spline_rep} is exact.}

{ \lemma \label{pchisq}

Let $\chi_j^2$ denote a $\chi^2$ random variable with degrees of freedom $j$. If $c\rightarrow \infty$ and $\frac{J}{c}\rightarrow 0$, then
$$P(\chi_j^2>c)=\frac{1}{\Gamma(j/2)}(c/2)^{j/2-1}\exp (-c/2)(1+o(1))$$
uniformly for all $j\leq J$.
}

The proof can be found in \cite{Luo-Chen13} by using integration by parts.

{ \lemma \label{chisq}

Let $\chi_j^2$ be a chi-square random variable with degrees of freedom $j$ and $c_j=2j[\log p+\log
(j\log p)]$. If $p\rightarrow \infty $, then for any $J\leq p$ and $h \ge 1$,
$$
\sum_{j=1}^J\binom{p}{j}P(\chi_{hj}^2>c_{hj})\rightarrow 0.
$$}

\begin{proof}
%This proof is a modified version of proof of Lemma~2 in \cite{Luo-Chen13}.
Let $q_j=\sqrt{\frac{c_{j}}{(j\log p)^2}}$.  By using $\binom{p}{j}\leq p^j$ and Lemma~\ref{pchisq},
\begin{align*}
\binom{p}{j}P(\chi_{hj}^2>c_{hj}) & = \binom{p}{j}\frac{1}{2^{hj/2-1}\Gamma(hj/2)}c_{hj}^
{hj/2-1}\exp(-c_{hj}/2) (1+o(1)) \\
& \le \frac{c_{hj}^{hj/2-1}} {{(hj \log p)}^{hj}} (1+o(1))\\
& = \frac{q_{hj}^{hj}}{c_{hj}} (1 + o(1))
\end{align*}
uniformly over $j < h J$ for any $J\le p$.

Since $q_j < 1$ for all $j$ and $q_j\rightarrow 0$ when $j$ is large enough, we have
$$\sum_{j=1}^J\binom{p}{j}P(\chi_{hj}^2>c_{hj})\leq \sum_{j=1}^J  \frac{q_{hj}^{hj}}{c_{hj}}
(1 + o(1))\rightarrow
0. $$
\end{proof}

\iffalse
{ \lemma \label{chisq2}

Let $\chi_j^2$ be a chi-square random variable with degrees of freedom $j$ and $c_j=2j[\log p+\log
(j\log p)]$. If $p\rightarrow \infty $, then for any $J\leq p$ and $h = o(\log(p))$,
$$
\sum_{j=1}^J\binom{p}{j}P(\chi_{hj}^2>c_{j})\rightarrow 0.
$$}
\fi

\subsection{Proof of Theorem~\ref{thm:1}}
Since
$$r(M) \propto (2 \pi)^{(p^*-n)/2}  2^{(n-p^*-2)/2} \RSS^{(p^*-n + 1)/2} \Gamma
\left(\frac{n-p^*}{2}\right) \times q^{p^*},$$
we have
$$\frac{r(M)}{r(M_0)}=\exp{\{-T_1-T_2\}},$$
where
$$T_1=\frac{n-h_nm-1}{2}\log\left(\frac{\RSS_M}{\RSS_{M_0}}\right)$$ and
$$T_2=\frac{h_n(m_0-m)}{2}\log{(\pi \RSS_{M_o})}+\log{\left\{\Gamma(\frac{n-h_nm_0}{2})/\Gamma(\frac{n-h_nm}{2})\right\}}+h_n(m_0-m)\log(q).$$

\noindent{\bf{Case 1}: $M_0\notin M$.}

Let $\mathcal{M}_j=\{M:|M|=j, M\in \mathcal{M}\}$.  Recall $\bH_M$ is the projection matrix for model $M$ and $\bH_{M_0}$ is the projection matrix for the true model $M_0$.  Calculate
\begin{align*}
 \RSS_{M_0}&=(\by-\bZ_{M_0}\bbeta_{M_0})^T(\bI-\bH_{M_0})(\by-\bZ_{M_0}\bbeta_{M_0}) \\
&=\besp^T(\bI-\bH_{M_0})\besp\\
 &=\sum_{i=1}^{n-h_nm_0}Z_i^2=(n-h_nm_0)(1+o_p(1))=n(1+o_p(1)),
\end{align*}
where $Z_i$'s are i.i.d. standard normal variables.

Let $\Delta (M)=\|\bmu-\bH_M\bmu\|$ with $\bmu=\bZ_{M_0}\bbeta_{M_0}$. Then
\begin{align} \label{RSS_diff}
\RSS_{M}-\RSS_{M_0}&=(\bmu+\besp)^T(\bI-\bH_M)(\bmu+\besp)-\besp^T(\bI-\bH_{M_0})\besp \nonumber  \\
&=\Delta(M)+2\bmu^T(\bI-\bH_M)\besp-\besp^T\bH_M\besp+\besp^T\bH_{M_0}\besp,
\end{align}
where $\besp^T\bH_{M_0}\besp=h_nm_0(1+o_p(1))$.

Express the second term in (\ref{RSS_diff}) as $$\bmu^T(\bI-\bH_M)\besp=\sqrt{\Delta(M)}Z_M,$$
where $Z_M\sim N(0,1)$.  Then for any $M \in \mathcal{M}$,
$$|\bmu^T(\bI-\bH_M)\besp|\leq \sqrt{\Delta(M)}\max_\mathcal{M}|Z_M|.$$

Let $c_j=2j\{\log p+\log (j\log p)\}$, according to Lemma~\ref{chisq} we have
\begin{align*}
P(\max_\mathcal{M}|Z_M|\geq \sqrt{c}) &= P(\max_{M\in\mathcal{M}_j,1\leq j\leq km_0}|Z_M|\geq \sqrt{c}) \\
&\leq\sum_{j=1}^{km_0}\binom{p}{j}P(\chi_1^2\geq c)\\
&\leq \sum_{j=1}^{km_0}\binom{p}{j}P(\chi_j^2\geq c)\rightarrow 0.
\end{align*}
Therefore, $|\bmu^T(\bI-\bH_M)\besp|=\sqrt{\Delta(M)O_p(km_o\log p)}$ uniformly over $\mathcal{M}$.

Similarly, for the third term in (\ref{RSS_diff}), as $\besp^T\bH_M\besp=\chi_{h_nm}^2$ we have
\begin{align*}
P(\max_\mathcal{M} \besp^T\bH_M\besp \geq c_{h_nj}) & =P(\max_{M\in\mathcal{M}_j,1\leq j\leq km_0} \chi_{h_nj}^2 \geq c_{h_nj}) \\
& \leq\sum_{j=1}^{km_0}\binom{p}{j}P(\chi_{h_nj}^2\geq c_{h_nj})\rightarrow 0.
\end{align*}
Thus we have
$$\max_{\mathcal{M}}\{\besp^T\bH_M\besp\}=O_p(kh_nm_o\log p).$$
Assuming that $h_nm_o\log p=o(n)$, we have
$$
\RSS_{M}-\RSS_{M_0}=\Delta(M)(1+o_p(1))
$$ and
\begin{align} \label{case1t1}
T_1&=\frac{n-h_nm-1}{2}\log\left(1+\frac{\RSS_M-\RSS_{M_0}}{\RSS_{M_0}}\right) \nonumber \\
&=\frac{n(1+o_p(1))}{2}\log\left\{1+\frac{\Delta(M)(1+o_p(1))}{n}\right\} \nonumber \\
& =\frac{\Delta(M)(1+o_p(1))}{2}.
\end{align}
By Sterling's formula,
$$
\log{\left\{\Gamma(\frac{n-h_nm_0}{2})/\Gamma(\frac{n-h_nm}{2})\right\}}=\frac{h_n (m-m_0)}{2}\log n(1+o(1)).
$$
Therefore
\begin{align} \label{case1t2}
T_2&=\frac{h_n (m-m_0)}{2}\left\{\log n (o_p(1)) -\log (\pi q^2)\right\} \nonumber \\
&\geq -\frac{h_nm_0}{2}\left\{\log n (o_p(1))-\log (\pi q^2)\right\}.
\end{align}

\noindent{\bf{Case 2}: $M_0\in M$.}

Let $\mathcal{M}^*$ be the collection of models that contain the true model; i.e., $\mathcal{M}^*=\{M\in\mathcal{M}, M_0\in M, M\neq M_0\}$. Moreover, let $\mathcal{M}^*_j=\{M,|M|=j,M_0\in M\}$.

When $M_0\in M$, $(\bI-\bH_M)\bZ_{M_0}=0$, therefore $\by^T(\bI-\bH_M)\by=\besp^T(\bI-\bH_M)\besp$.  Also
\begin{align*}
\RSS_{M}-\RSS_{M_0} & =\besp^T(\bI-\bH_{M_0})\besp-\besp^T(\bI-\bH_M)\besp \\
& =\besp^T(\bH_M-\bH_{M_0})\besp \\
& =\chi_{h_n(m-m_0)}^2(M),
\end{align*}
where $\chi_{h_n(m-m_0)}^2(M)$ follows chi-square distribution with degrees of freedom $h_n(m-m_0)$.

Let $c_j=2j\{\log p+\log (j\log p)\}$. According to Lemma~\ref{chisq},
\begin{align*}
P(\max_{M\in\mathcal{M}^*_j,1\leq j\leq km_0-m_0} \chi_{h_nj}^2(M) \geq c_{h_nj})&=\sum_{j=1}^{km_0-m_0}P(\max_{M\in\mathcal{M}^*_j} \chi_{h_nj}^2(M) \geq c_{h_nj}) \\ &=\sum_{j=1}^{km_0-m_0}\binom{p-m_0}{j} P( \chi_{h_nj}^2(M) \geq c_{h_nj}) \\
&=\sum_{j=1}^{km_0-m_0}\binom{p}{j} P( \chi_{h_nj}^2(M) \geq c_{h_nj}) \rightarrow 0.
\end{align*}
Therefore, $\chi_{h_n(m-m_0)}^2(M)\leq c_{h_n(m-m_0)}(1+o_p (1))$ and
\begin{align*}
T_1&=\frac{n-h_nm-1}{2}\log\left(\frac{\RSS_M}{\RSS_{M_0}}\right) \\
&=-\frac{n-h_nm-1}{2}\log\left\{1+\frac{\chi_{h_n(m-m_0)}^2(M)}{\RSS_{M_0}-\chi_{h_n(m-m_0)}^2(M)}\right\}\\
&\geq -\frac{n-h_nm-1}{2}\left\{\frac{\chi_{h_n(m-m_0)}^2(M)}{\RSS_{M_0}-\chi_{h_n(m-m_0)}^2(M)}\right\}.
\end{align*}
Since $n^{-1}\RSS_{M_0}\rightarrow \sigma^2$ as $n\rightarrow \infty$, we have $\RSS_{M_0}=n(1+o(1))$,
\begin{align} \label{case2t1}
T_1&\geq \frac{c_{h_n(m-m_0)}}{2}(1+o_p(1)) \nonumber \\
&\geq -h_n(m-m_0)\left[1+\frac{\log\{h_n(km_0-m_0)\log p\}}{\log p}\right]\log p(1+o_p(1)) \nonumber \\
&\geq -h_n(m-m_0)(1+\delta)\log p(1+o_p(1))
\end{align}
uniformly over $\mathcal{M^*}$, and
\begin{align} \label{case2t2}
T_2=\frac{h_n (m-m_0)}{2}\{\log n (o_p(1))
-\log (\pi q^2)\}
\end{align}
uniformly over $\mathcal{M^*}$.

Combing case $1$ and case $2$, we aim to show that
$$\max_{M\notin M_0, M\in \mathcal{M}}\frac{r(M)}{r(M_0)}
=\max\{\max_{M_0\notin M}\exp (-T_1-T_2), \max_{M_0\in M}\exp (-T_1-T_2)\}\rightarrow 0. $$

By (\ref{case1t1}) and (\ref{case1t2}), for case $1$,
\begin{align*}
T_1+T_2 &\geq \frac{\Delta(M)(1+o_p(1))}{2} - \frac{h_nm_0}{2}\left\{\log n (o_p(1)) -\log (\pi q^2)\right\} \\
&=\frac{h_nm_0\log p}{2}\left\{ \frac{\Delta(M)(1+o_p(1))}{h_nm_0\log p}-\frac{\log n o_p(1)}{\log p} +\frac{\log (\pi q^2)}{\log p} \right\}.
\end{align*}
In order that
$$\min_{M_0\notin M}T_1+T_2 \rightarrow \infty,$$
we can choose $q$ such that $-\log q=O(\log p)$; i.e.
$$-\frac{\log q}{\log p}=O(1).$$

Similarly by (\ref{case2t1}) and (\ref{case2t2}), for case $2$,
\begin{align*}
T_1+T_2
&\geq \frac{h_n(m-m_0)\log p}{2}\left\{ \frac{\log n (o_p(1))}{\log p} -\frac{\log (\pi q^2)}{\log p}-2(1+\delta)(1+o_p(1)) \right\}.
\end{align*}
In order that
$$\min_{M_0\in M}T_1+T_2 \rightarrow \infty,$$
we have
$$-\frac{\log q}{\log p}>1+\delta.$$
Therefore, for $1+\delta<\gamma=-\frac{\log q}{\log p}<C$ with $C$ being a constant, we have
$$\max_{M\notin M_0, M\in \mathcal{M}}\frac{r(M)}{r(M_0)}\rightarrow 0.$$

Moreover, if condition (A7) holds, we have
\begin{align*}
\sum_{M\neq M_0,M\in \mathcal{M}^*}\frac{r(M)}{r(M_0)}
& \leq \sum_{j=1}^{km_0}\sum_{\mathcal{M}^*}\frac{r(M)}{r(M_0)} \\
& \leq km_0\max_{M\neq M_0,M\in\mathcal{M}}|M_j^*|\frac{r(M)}{r(M_0)}\rightarrow 0.
\end{align*}
This completes the proof for Theorem~\ref{thm:1}.

\bibliography{references}
\bibliographystyle{rss}

\end{document}